\documentclass[11pt,a4paper,twocolumn]{article}
\usepackage[T1]{fontenc}
\usepackage[utf8]{inputenc}
\usepackage{mathpazo}          
\usepackage[a4paper,top=20mm,bottom=20mm,left=20mm,right=20mm]{geometry}
\usepackage{amsmath,amssymb}
\usepackage{booktabs}
\usepackage{array}
\usepackage{microtype}          
\usepackage[hidelinks]{hyperref}

\newcommand{\subhead}[1]{%
  \par\addvspace{9pt}%
  {\bfseries\large #1\par}%
  \addvspace{2pt}\noindent\ignorespaces}

\newcommand{\sem}[1]{\mathopen{[\mkern-1.5mu[}#1\mathclose{]\mkern-1.5mu]}}
\newcommand{\Out}{\mathit{Out}}

\begin{document}
\thispagestyle{empty}
\pagestyle{empty}

\twocolumn[{%
\begin{center}
{\fontsize{27}{31}\selectfont\bfseries The Café in Amsterdam}\\[5pt]
{\Large\itshape When the Incumbent Becomes the Oracle}\\[8pt]
{\footnotesize Augusto Camargo \ \textbullet\ \ Bluecore Consulting, São Paulo, Brazil \ \textbullet\ \ augusto.camargo@bluecore.com.br \ \textbullet\ \ 25 August 2026}\\[10pt]
\end{center}
\noindent\rule{\textwidth}{0.4pt}\\[10pt]
{\centering
\begin{minipage}{0.66\textwidth}
\centering\small\itshape
A research note. It proposes a way of looking at computational reformulation --- a lens, not a theorem.
No component is claimed novel in isolation; the offering is the synthesis, and the question it makes easy
to ask.
\end{minipage}\par}%
\vspace{10pt}%
}]

In 1956, sitting in a café in Amsterdam, Edsger Dijkstra worked out---in about twenty
minutes, with no pencil---how a computer should find the shortest route between two Dutch cities. He
needed a concrete problem to show off the ARMAC machine, so he picked one anybody could check on a map:
the shortest way from Rotterdam to Groningen. The algorithm that came out of that afternoon is now taught
to virtually every computer science student on Earth~\cite{dijkstra-interview}.

Notice the order in which he did it, because the whole idea I want to offer lives inside that order.

\textbf{He formulated the demand first.} \emph{Find a path of minimum total length.} That sentence exists
independently of any algorithm---it is a specification, not an implementation. The algorithm came second,
as one way, out of many, to satisfy it. Dijkstra never confused the two. The demand was \emph{a shortest
path}. His procedure was \emph{one method for producing one}.

And because the demand existed independently of the algorithm, the field stayed free to reinvent the
algorithm entirely. Over the following sixty years, shortest-path computation was reformulated almost
beyond recognition: A\textsuperscript{*} adds a heuristic; Contraction Hierarchies precompute shortcuts
and answer continental queries in microseconds; hub labelling, transit-node routing, bidirectional
search, landmark methods---a family whose members share remarkably little internal structure.

\textbf{Not one of them was ever required to reproduce Dijkstra's path.} A new algorithm is not asked
whether it visits the same intermediate cities. It is asked whether it satisfies the same specification:
return a shortest path---or, in the approximate variants, one whose gap from shortest is explicitly
declared. Establishing that is not trivial: summing a path's weights tells you its cost, not its
optimality. But the algorithms can carry their own evidence---distance labels and a predecessor tree form
a certificate checkable in time linear in the graph. So the freedom never depended on verification being
free. It depended on this: \textbf{the obligation was stated without mentioning the incumbent implementation.} From this point on, I
simply call it the incumbent. Dijkstra's
algorithm was one solver of a named problem. It was never the definition of a correct answer.

\subhead{The stakes are silicon and joules}

This is not an antiquarian point. It seems to decide what a field can afford to build. Modern
accelerators pay for one shape of computation---dense, regular, reused---and a problem posed in that shape
can run several times faster, at a fraction of the energy, than the same problem posed the way a
sequential machine liked it. When you are free to change the formulation, that speedup is on the table.
When you are not, it is not. A speech frontend we re-posed as a single matrix multiplication measured
$1.64\times$--$3.29\times$ speedup and up to $3.03\times$ less energy across four accelerators---not by
computing the incumbent's answer faster, but by computing a \emph{different} answer that the task could
not distinguish from it~\cite{melt}. Whether a field is allowed to reach for that trade is, it seems to
me, exactly the question the café quietly decided.

\subhead{A computation whose café conversation never happened}

The dominant audio frontend in speech processing is a pipeline: a Fourier transform, a Mel filterbank, a
logarithm. Decades of engineering have tuned it, vendors have accelerated it, and accelerators still
inherit its assumptions. Ask the awkward question---\emph{what is the implementation-independent demand
that a structurally different frontend must satisfy to replace it?}---and the answer is not that audio is
blind to it. \textbf{Audio, of all fields, has the oracle.} SincNet and LEAF replace the Fourier--Mel
frontend with entirely different objects and are admitted purely on downstream task
accuracy~\cite{sincnet,leaf}; the community plainly accepts a task-level criterion when it wants one.

But look at where that criterion was pointed. It was pointed at \emph{representation}---at learning a
task-adapted frontend---and, as far as I can see, never at the \emph{substrate}. When those systems
changed the formulation, they changed it in the direction that made it \emph{slower}: LEAF is far more
expensive to compute than a fixed Mel filterbank, and its own successors report that it fails to
consistently beat one~\cite{efficientleaf}. Meanwhile the part of the field that \emph{did} care about
speed---the ports, the vendor kernels---judged a replacement by whether it reproduced the pipeline's
output. So the demand-side oracle existed and the substrate motive existed, in the same field, and---this
is the thing I keep turning over---I had not seen the two put together. Where the substrate was the reason
to change, the incumbent's output was still the test.

That is the phenomenon I want to point at. Not a field with no ruler---a field holding a ruler it never
thought to use for this. The consequence is the same either way: a genuinely different frontend does not
lose on the merits. On the only ruler in routine use, \emph{proximity to the incumbent's output}, it
cannot be scored at all. It looks wrong because it is different, even when it serves the task equally
well.

When a field has an implementation-independent demand, the incumbent looks like mere \emph{evidence} that
the demand can be met. When it lacks one---or never aims the one it has---the incumbent itself seems to
become the practical oracle. I have taken to calling that transition \textbf{baseline capture}: the
moment an incumbent stops being evidence and becomes judge. The incumbent only ever proved the demand
\emph{could} be satisfied. It never proved the demand \emph{had} to be satisfied that way.

\begin{table*}[t]
\begingroup\centering\small\setlength{\tabcolsep}{5pt}
\begin{tabular}{@{}>{\raggedright\arraybackslash}p{3.4cm}>{\raggedright\arraybackslash}p{4.0cm}>{\raggedright\arraybackslash}p{3.7cm}>{\raggedright\arraybackslash}p{3.6cm}@{}}
\toprule
& \textbf{demand stated independently of the incumbent?} & \textbf{how a candidate is justified} & \textbf{what happened} \\
\midrule
Routing \newline (Dijkstra $\to$ A\textsuperscript{*} $\to$ CH) &
\textbf{yes} --- shortest path, or a declared approximation &
proof, certificate, or bound against the \emph{stated objective} &
continuous reformulation for six decades \\
\addlinespace[3pt]
Audio frontend, \newline for the substrate &
\textbf{available, not applied here} --- the task oracle exists, yet substrate-motivated replacement is judged by matching the pipeline &
agreement with the established representation &
deep optimization \emph{within} the incumbent family \\
\bottomrule
\end{tabular}\par\endgroup
\end{table*}

\subhead{The lens, in four lines}

The prose above wants only a little notation, and no more than this. Let $\mathbb{O}$ be the set of
possible outputs, and write $\Out \models D$ when an output satisfies the demand $D$. Then
\[
\begin{aligned}
&\text{$P$ represents $D$:}   & & \exists\, \Out \in \sem{P}.\ \Out \models D;\\
&\text{$P$ is sound for $D$:} & & \forall\, \Out \in \sem{P}.\ \Out \models D.
\end{aligned}
\]
An acceptance test is a predicate $T:\mathbb{O}\to\{0,1\}$, and the whole distinction is \emph{what
$T$ is built from}:
\[
\begin{aligned}
T_{D}(\Out) &= \mathbf{1}[\,\Out \models D\,] \quad\text{\small(independent)}\\[2pt]
T_{0}(\Out) &= \mathbf{1}[\,\Out \in R(\Out_0)\,] \quad\text{\small(captured)}
\end{aligned}
\]
where $R(\Out_0)$ is any acceptance region defined from the incumbent's \emph{output} rather than from
$D$; its extreme case $R(\Out_0)=\{\Out_0\}$ is exact regression. \emph{Baseline capture} is the drift
$T_D \rightsquigarrow T_0$. Writing $A(T)=\{P : \exists\,\Out\in\sem{P},\ T(\Out)=1\}$ for the
representations a test admits, $A(T_D)$ is everything that can serve the demand, while $A(T_0)$ is only
what can reach the incumbent. Nothing here needs a metric, an $\varepsilon$, or a margin. The one question a reader can
carry to any paper is just: \emph{does the definition of $T$ mention $\Out_0$?}

Two properties of a $\langle\text{problem},\text{field}\rangle$ then pull apart, and should not be
confused. Whether a reformulation can be \emph{judged at all} looks like it turns on whether an
incumbent-independent $T_D$ exists. Whether its discovery can be \emph{automated}---dropped inside a
program-search loop, as FunSearch or AlphaEvolve do~\cite{funsearch,alphaevolve}---looks like it turns
additionally on whether $T$ is cheap to evaluate. Routing has both: an independent objective and a
linear-time certificate. The matrix-multiplication frontend has the first but not the second---a frozen
model is an independent $T_D$, but it costs hours to consult, which may be exactly why finding the
reformulation stayed a human, experimental act. In the captured case the missing ingredient does not look
primarily like a cheap evaluator. It looks like a shared, incumbent-independent demand.

\subhead{Two honesties}

\textbf{The mechanism is old. The question I want to ask about it is simple, and it may not be mine
alone.} Software testing has lived with the oracle problem for decades. Weyuker described
\emph{non-testable programs}, whose correct output cannot be independently determined~\cite{weyuker}. Barr
and colleagues surveyed the substitutes the field reaches for---pseudo-oracles, previous versions, golden
outputs---and noted that such a substitute can ``migrate, over time, to become'' the
specification~\cite{barr}. Requirements engineering calls a close cousin \emph{implementation bias}:
stating \emph{how} instead of \emph{what}~\cite{zave}. I lean on all of it, gratefully---that serious
people have named these pieces is what makes me think the lens is real rather than a private illusion.
The question I keep wanting to ask, and have not often seen asked in these words, is what this
substitution \emph{costs a computation}: whether the practice quietly decides which reformulations a
field is willing, or even able, to admit---and therefore how fast that field is allowed to run.

\textbf{Judging by acceptability rather than numerical proximity alone is not new.} Felzmann, Filho, de
Oliveira and Wanner argued for evaluating approximate designs in terms of whether their outputs remain
acceptable for the application, rather than treating conventional quality metrics as
sufficient~\cite{felzmann}. Theirs is prior work and I am glad to stand next to it. The distinction I
find myself caring about is narrower: approximate computing ordinarily begins from an incumbent
computation and asks which deviations from it remain acceptable. The cases that made me look twice need
not be naturally described as degraded versions of that computation. A Mel-spaced projection is not
simply an inaccurate FFT; there is no obvious parameter that continuously turns one into the other. So
the question I am left with is not \emph{how much quality may I spend against the incumbent}, but
\emph{when is a structurally different formulation admissible at all}---and the incumbent's output,
alone, does not seem able to answer that.

\subhead{What buying a verifier looks like}

The escape, if it is one, is not cleverness. It looks more like bookkeeping done late. When a community
finally makes its acceptance condition explicit, reformulations that had been unjustifiable for years
seem to become admissible almost at once. \textbf{ZIP~215:} independent implementations of Ed25519 had
inherited their notion of a valid signature from an obsolete build of a library---bug and all---and could
disagree on which signatures were valid; the fix was to \emph{write the validation rules down
explicitly}, which let implementations agree and, as a named reward, enabled batch
validation~\cite{zip215}. \textbf{CESM-ECT:} climate simulations cannot be compared bit-for-bit across
compilers, so instead of handing four hundred simulated years to a senior scientist for subjective
approval, the community built a statistical test asking whether a run represents \emph{the same climate}
rather than the same bits~\cite{cesmect}. In each case someone made the acceptance condition explicit,
operational, and independent of the incumbent. That is all I mean by \emph{buying a verifier}: not
necessarily writing an objective function---CESM-ECT is a consistency test, not an objective---but making
the demand answerable without pointing back at the incumbent. It has a price, and the guess I would like
to test is that the price is measurable, and that paying it enlarges the space of reformulations, and
thus the hardware performance, a field can reach.

\subhead{Coda}

Dijkstra spent twenty minutes in a café and, without meaning to, handed an entire field sixty years of
freedom. Most fields never had that conversation. Whether this phenomenon deserves a name is ultimately
less important than whether it exists. If it does, then making a field's demand explicit---stating it
without mentioning the incumbent---may be one of the cheapest ways we have of enlarging the space of
admissible reformulations, and one of the few that can be done before a single line of the new
computation is written.

\onecolumn

\end{document}